\documentclass[final,7p,times,twocolumn]{elsarticle}

\begin{document}

\begin{frontmatter}

\title{Anisotropic magnetization, resistivity and heat capacity of single crystalline \textit{R}$_3$Ni$_{2-x}$Sn$_7$ (\textit{R} = La, Ce, Pr and Nd)}

\author[label1]{Xiao Lin}  \author[label1]{Sergey L. Bud'ko}  \author[label2]{Srinivasa Thimmaiah}  \author[label1]{Paul C. Canfield}

\address[label1]{Department of Physics and Astronomy and Ames Laboratory, Iowa State University, Ames, Iowa 50011, U.S.A.}

\address[label2]{Department of Chemistry and Ames Laboratory, Iowa State University, Ames, Iowa 50011, U.S.A.}

\begin{abstract}
We present a detailed study of \textit{R}$_3$Ni$_{2-x}$Sn$_7$ (\textit{R} = La, Ce, Pr and Nd) single crystals by measurements of crystal structure, stoichiometry, temperature dependent magnetic susceptibility, magnetization, electrical resistivity, magnetoresistance, and specific heat. This series forms with partial Ni occupancy with \textit{x} varying from $\sim$ 0.1 for \textit{R} = La to $\sim$ 0.7 for \textit{R} = Nd. The electrical resistivity of this series follows metallic behavior at high temperatures. Determination of clear anisotropies as well as antiferromagnetic ordering temperatures for \textit{R}$_3$Ni$_{2-x}$Sn$_7$ (\textit{R} =Ce, Pr and Nd) have been made. For Pr$_3$Ni$_{1.56}$Sn$_7$ and Nd$_3$Ni$_{1.34}$Sn$_7$, multiple magnetic transitions take place upon cooling. Metamagnetic transitions in this family (\textit{R} = Ce, Pr and Nd) were detected for applied magnetic fields below 70 kOe. An $H-T$ phase diagram of Ce$_3$Ni$_{1.69}$Sn$_7$ was assembled to shed light on its low field properties and to rule out possible quantum critical effects.

\end{abstract}


\end{frontmatter}

\section{Introduction}

Rare earth (\textit{R}) compounds have always been of great interest to experimentalists and theorists for their various unusual magnetic, electronic and structural properties \cite{Handbook, Handbook2, Sergey_1999, Sefat_2008, Mun_2010, Sergey_2000}. For the majority of rare-earth elements and compounds, the 4\textit{f}  electrons are shielded from the 5\textit{s}-, 5\textit{p}- and 4\textit{d}-shell electrons, and thus do not participate in chemical bonding. However, since the magnetic moments are from the 4\textit{f} electrons, the \textit{R}-bearing compounds can manifest vastly different magnetic properties. By varying the \textit{R} elements in a compound, it is possible to tune the magnetism and other physical properties. Known as the lanthanide contraction, the unit cell volume of isostructural $R^{3+}$-bearing families shrink across the series. This contraction can lead to systematic changes in the lattice constants \textit{a, b, c}, and maybe eventually drive the series out of its structural stability. In addition, the crystalline electric field (CEF) also plays a significant role in determining the temperature-dependent thermodynamic and transport properties of a compound. Associated with the point symmetry of the \textit{R} ions, the CEF splitting can cause anisotropy, and influence the spin arrangements in the ordered state, often affecting the details of metamagnetic transitions. In terms of the ordering temperature and energy scales, the CEF splitting affects the amount of entropy (associated with the 4\textit{f} electrons) that can be removed. Thus, a comparative study of a series of rare-earth compounds can give an insight into the evolution of the rich and complex physics. Here, we present a study of physical properties of single crystalline samples of the \textit{R}$_3$Ni$_{2-x}$Sn$_7$ (\textit{R} = La, Ce, Pr and Nd) series.

The early report of \textit{R}$_3$Ni$_{2-x}$Sn$_7$ can be traced back to late 1980s \cite{Skolozdra_1987}. The structure was solved based on polycrystalline samples of \textit{R}$_3$Ni$_2$Sn$_7$ (\textit{R} = La, Ce, Pr and Nd). Based on neutron diffraction \cite{SP_2001}, Ni and Sn sites were reported to have partial occupancies. Although magnetic susceptibility, electrical resistivity and thermopower of this series were measured between 78-350 K, practically no low-temperature properties were reported for \textit{R} = La, Pr and Nd.

Later studies were focused on Ce$_3$Ni$_2$Sn$_7$ polycrystalline samples \cite{SP_2001,Chevalier_1999,Chevalier_2001,Matar_2003}. The ground state of Ce-based intermetallic compounds is often governed by the competition between the Ruderman-Kittel-Kasuya-Yosida (RKKY) interaction and the Kondo interaction. Depending on the strength of the hybridization between 4\textit{f} and conduction electrons relative to their coupling strength, the ground state can be either a non-magnetic state dominated by the Kondo interaction or a long-range magneticly ordered state governed by the RKKY interaction. Ce$_3$Ni$_2$Sn$_7$ is reported to order antiferromagneticly at $T_{\rm N}$ $\sim$ 3.8 K \cite{SP_2001, Chevalier_1999}. Ce$_3$Ni$_2$Sn$_7$ crystallizes in the orthorhombic structure ($Cmmm$, No. 65) \cite{Skolozdra_1987, SP_2001} and Ce occupies two different crystallographic sites (the Ce1 \textit{2c} site: \textit{mmm} and the Ce2 \textit{4i} site: \textit{m2m}), but studies have shown that only the Ce2 atoms with a trigonal prism arrangement participate in the magnetic ordering \cite{SP_2001}. Moreover, based on measurements on polycrystalline Ce$_3$Ni$_2$Sn$_7$, metamagnetic transitions at low temperatures were inferred, indicating a complex spin arrangement \cite{Chevalier_2001}. Since metamagnetism in rare earth compounds is usually very anisotropic, single crystals are required for systematic  studies.

In this paper, we present a systematic study of the anisotropic properties of the \textit{R}$_3$Ni$_{2-x}$Sn$_7$ series with \textit{R} = La, Ce, Pr and Nd. Since this system shows partial occupancy of the Ni site, a detailed structural study and refinement of the site occupancies are also provided in this work. Measurements and analyses of the field and temperature dependence of magnetization, resistivity and specific heat were performed on single crystalline samples. Measurements of the magnetization parallel to the \textit{b}-axis and the \textit{ac}-plane show anisotropic behavior, and the magnetization of some of the compounds manifest metamagnetic transitions. 

\begin{figure}
\begin{center}
\resizebox*{7.5cm}{!}{\includegraphics{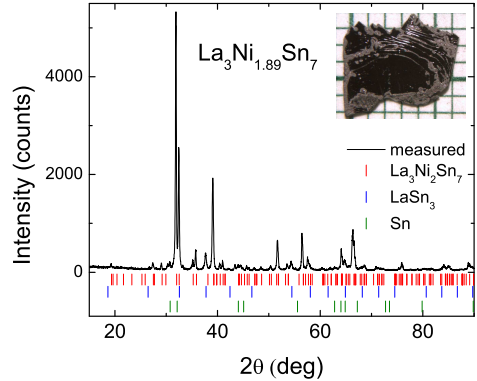}}%
\caption{ Powder x-ray diffraction pattern of La$_3$Ni$_{1.89}$Sn$_{7}$. Inset: picture of single crystalline La$_3$Ni$_{1.89}$Sn$_{7}$ on a millimeter grid.}%
\label{fig:x-ray}
\end{center}
\end{figure}

\section{Experimental details}
Single crystals of \textit{R}$_3$Ni$_{2-x}$Sn$_7$ (\textit{R} = La, Ce, Pr and Nd) were grown out of excess Sn flux via the high-temperature solution method \cite{Canfield_1992}. High purity elements ($>$3N), with an initial composition of 9:18:73 (\textit{R}:Ni:Sn) were used in the synthesis. The constituent elements were placed in an alumina crucible and sealed in a silica tube under a partial pressure of high purity argon gas. This was then heated up to 1100$^\circ$C and slowly cooled to 800$^\circ$C, at which temperature the excess solution was decanted using a centrifuge. Single crystals of \textit{R}$_3$Ni$_2$Sn$_7$ grew in plate-like shape with their largest dimensions limited to the size of the crucible (see inset of Fig. \ref{fig:x-ray}). The crystallographic \textit{b}-axis is perpendicular to the plane of the plate-like single crystals. Most of the samples had shiny surfaces that were partially covered by secondary phase materials. Since HCl was found to attack the surface material as well as \textit{R}$_3$Ni$_{2-x}$Sn$_7$, and due to Sn's malleable nature, the samples were not etched or polished in the following measurements. Although all samples in this study were grown from the same initial stoichiometry, a significant and variable Ni deficiency (\textit{R}$_3$Ni$_{2-x}$Sn$_7$) develops as \textit{R} changes from La to Nd. 

Powder x-ray diffraction data were collected on a Rigaku MiniFlex diffractometer with Cu K$\alpha$ radiation at room temperature. The error bars were determined by statistical errors, and standard Si powder was used as the internal reference.
 
Single crystal x-ray diffraction data were collected by Bruker APEX diffractometer equipped with a CCD detector, using monochromated Mo K$\alpha$ radiation ($\lambda$ = 0.71073 \r{A}). Reflections were gathered by taking three sets of 606 frames with 0.3$^{\circ}$ scans in $\omega$ and with an exposure time of 10 \textit{s} per frame at room temperature. The range of 2$\theta$ extended from $\sim$ 4 to 57$^{\circ}$. The reflection intensities were integrated with the \textit{SAINT} program. The measured intensities were corrected for Lorentz and polarization effects and were further corrected for absorption using the \textit{SADABS} program as implemented in the \textit{SAINT} \cite{X1} program package. Intensity statistics and space group determination were carried out using \textit{XPREP}, a subprogram in the \textit{SHELXTL} software package. The structural models were obtained from direct methods using \textit{SHELXS}-­97 and refined by full­matrix, least­squares procedures on $F^2$ as implemented in the \textit{SHELXTL} \cite{X2} package. 

Elemental analysis of the samples was performed using wavelength dispersive x-ray spectroscopy (WDS) in the electron probe microanalyzer of a JEOL JXA-8200 Superprobe. For each compound, the WDS data were collected from multiple locations on multiple samples. To determine the bulk concentration, only clear and shiny surface regions were selected for these measurement, i.e. regions with residual Sn flux were avoided. 

Measurements of field dependent magnetization and temperature dependent susceptibility were performed in a Quantum Design, Magnetic Property Measurement System (MPMS). The ac resistivity was measured by a standard four-probe method in a Quantum Design, Physical Property Measurement System (PPMS) and with LR700 ac resistance bridge in MPMS. Platinum wires were attached to the sample using either Dupont 4929 silver paint or Epotek H20E silver epoxy with the current flowing in the \textit{ac}-plane. As mentioned above, given the malleable nature of the flux and fragile samples, no polishing was done prior to the resistivity measurement. The room temperature resistivity of this series ranges from 50 - 100 $\mu\Omega$ cm. The absolute values of resistivity are accurate to $\pm$ 50$\%$ due to the irregularity of the sample geometry and positions of electrical contacts. For data presentation, the basal plane resistivity is normalized with respect to the room temperature value assuming that current density was uniformly distributed throughout the cross section. The residual resistivity ratio is determined as (RRR) = $\rho$(300 K) / $\rho$(6.5 K) for La$_3$Ni$_{1.89}$Sn$_{7}$; and (RRR) = $\rho$(300 K) / $\rho$(1.8 K, 0.5 kOe) for \textit{R}$_3$Ni$_{2-x}$Sn$_7$ (\textit{R} = Ce, Pr and Nd). To remove the high frequency noise caused by digital differentiation of closely spaced data points, an FFT filter method provided by Origin 8.5 program was used in calculating the temperature and field derivatives of $\rho$ and \textit{M} \cite{Origin}.

A relaxation technique was used in the heat capacity measurements in the PPMS. The specific heat data of La$_3$Ni$_{1.89}$Sn$_7$ was used to estimate the non-magnetic contributions to the specific heat of \textit{R}$_3$Ni$_{2-x}$Sn$_7$ (\textit{R} = Ce, Pr and Nd). Here we assume the differences in non-magnetic specific heat brought by different Ni site deficiencies are negligible. The magnetic contribution to specific heat from the \textit{R} ions was calculated by the relation of $C_M$ = $C_p$ (\textit{R}$_3$Ni$_{2-x}$Sn$_7$) - $C_p$ (La$_3$Ni$_{1.89}$Sn$_7$). A linear extrapolation was used to estimate $C_M$'s behavior down to zero temperature. The magnetic entropy $S_M$ for \textit{R} = Ce, Pr and Nd members was calculated by integrating $C_M/T$ per mole \textit{R} with the measured and extrapolated data. 

\section{Results and analysis}
\subsection{Crystal stoichiometry and structure}

Powder x-ray diffraction patterns were collected on ground single crystals from each compound. Figure \ref{fig:x-ray} shows a La$_3$Ni$_{1.89}$Sn$_7$ x-ray pattern as an example. The main phase was resolved to be La$_3$Ni$_{1.89}$Sn$_7$, and small traces of Sn residue as well as LaSn$_3$ can be detected in the diffraction pattern. Similar results (\textit{R}$_3$Ni$_{2-x}$Sn$_{7}$ with minority phases of \textit{R}Sn$_3$ and Sn) were obtained for the other members of the series. The analysis of powder x-ray diffraction data indicates that the lattice parameters \textit{a}, \textit{b} and \textit{c} are monotonically decreasing as the series progresses from La to Nd (presented in Table \ref{table:powder}).

\begin{figure}
\begin{center}
\resizebox*{7.5cm}{!}{\includegraphics{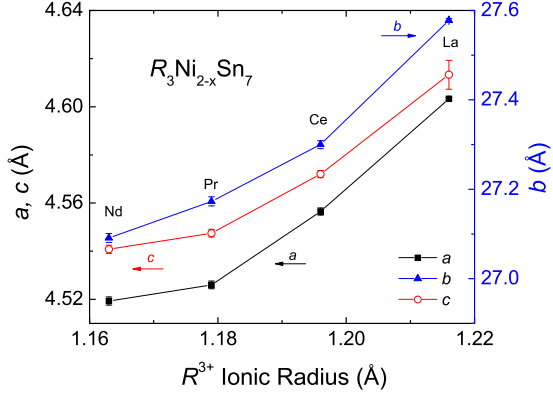}}%
\caption{ The change of unit cell lattice parameters vs. ionic radius of $ R^{3+}$ \cite{Shannon_1976} in \textit{R}$_3$Ni$_{2-x}$Sn$_7$ compounds, refined from single crystal x-ray diffraction data.}%
\label{fig:lattice}
\end{center}
\end{figure}

Since site occupancy was identified as a potential problem \cite{SP_2001}, room-temperature single crystal diffraction data were also collected. Table \ref{table:unit} summarizes the lattice constants and Ni site occupancies. The refined positional parameters for \textit{R}$_3$Ni$_{2-x}$Sn$_7$ (\textit{R}=La, Ce, Pr and Nd) series are included in Table \ref{table:occupancy}. Proceeding from the larger to the smaller rare-earth elements, all lattice parameters decrease almost linearly: 1.8$\%$ for \textit{a}; 1.8$\%$ for \textit{b} and 1.6$\%$ for \textit{c} (as shown in Fig. \ref{fig:lattice}), which is consistent with the results of powder x-ray analysis (Table \ref{table:powder}) and previously reported data \cite{Skolozdra_1987}. Furthermore, the overall volume decreases by ~5.1$\%$. These results are due to the lanthanide contraction that occurs across the 4\textit{f} series as well as the decreasing Ni occupancy. The ionic radius of trivalent rare-earth was taken from ref. \cite{Shannon_1976} for 9 coordination number (CN=9).

The stoichiometry of the \textit{R}$_3$Ni$_{2-x}$Sn$_7$ (\textit{R} = La, Ce, Pr and Nd) samples was also inferred from WDS analyses. The averaged atomic percentages of each element in each compound are normalized to $R_{3.00}$ (Table \ref{table:WDS}). The results show that although the ratio of \textit{R}:Ni:Sn is close to 3:2:7, significant Ni deficiency develops as the atomic number of rare earth elements increases. The stoichiometries inferred from the WDS and single crystal diffraction data are qualitatively similar (Table \ref{table:unit} and \ref{table:WDS}). Although there are slight quantitative differences, the clear trend in all increasing Ni deficiency from La to Nd is clear. 

Given this series of compounds does not maintain a fixed, stoichiometric composition for all rare-earth samples, the calculation of physical quantities, such as the magnetization and specific heat, the actual stoichiometries from Table \ref{table:unit} will be used.

\subsection{La$_3$Ni$_{1.89}$Sn$_7$}

\begin{figure}
\begin{center}
\resizebox*{7.5cm}{!}{\includegraphics{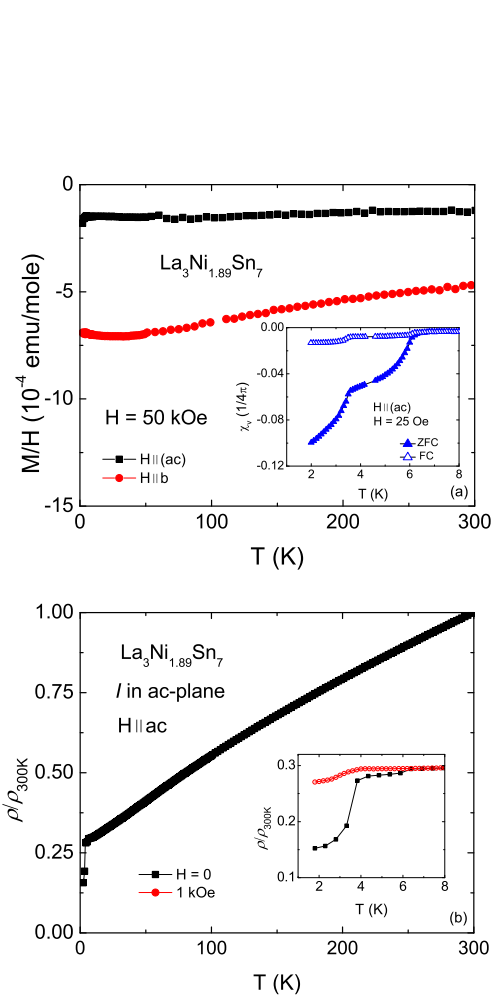}}%
\caption{ (a) Magnetic susceptibility of La$_3$Ni$_{1.89}$Sn$_7$ for \textit{H} = 50 kOe, with fields both parallel to \textit{b}-axis and \textit{ac}-plane. Inset: ZFC and FC magnetic susceptibility for \textit{H} = 25 Oe. (b) Temperature dependence of the normalized resistivity ratio of La$_3$Ni$_{1.89}$Sn$_7$ with $\rho$(300 K) $\sim$ 50 $\mu\Omega$ cm. Inset: enlarged normalized resistivity ratio for $T \leq$ 8 K in zero field and 1.0 kOe.}%
\label{fig:La_MR}
\end{center}
\end{figure}

The magnetic susceptibility of La$_3$Ni$_{1.89}$Sn$_7$ measured in an applied field of 50 kOe (Fig. \ref{fig:La_MR}(a)), is negative and exhibits an almost temperature independent behaviour from 2 K to 300 K. It also manifests a relatively large anisotropy between the magnetic field parallel to the \textit{b}-axis and the \textit{ac}-plane, with $\vert(M/H)_{b}\vert > \vert(M/H)_{ac}\vert$. Neither the diamagnetism nor anisotropy is uncommon, similar behaviors have been reported for other La-based compounds \cite{Sefat_2008, Myers_1999}.

The temperature dependence of the normalized zero-field resistivity ratio of La$_3$Ni$_{1.89}$Sn$_7$ (Fig. \ref{fig:La_MR}(b)) displays metallic behaviour with (RRR) $\simeq$ 3.2 for current in the \textit{ac}-plane. To within a factor of 20\%, the room temperature resistivity value $\rho$(300 K) reaches $\sim$ 50 $\mu\Omega$ cm. We assume that this small RRR is at least partially due to the deficiencies at the Ni site. At low temperatures, two resistive anomalies are observed near 6.2 K and 3.8 K (inset of Fig. \ref{fig:La_MR}(b)). When measured in an applied magnetic field of 1.0 kOe, the higher-temperature anomaly shifts to lower temperature and the lower one disappears. The ZFC and FC superconducting fractions are also estimated by magnetization measurement at 25 Oe with the field parallel to the \textit{ac}-plane, as shown in the inset of Fig. \ref{fig:La_MR}(a). The small values of FC superconducting fractions, $\sim$ 1.4$\%$, and ZFC fraction, as well, $<$ 15$\%$ indicate that superconductivity is filamentary and the anomalies in the resistivity can be attributed to impurities. In fact, it is highly likely that these two anomalies are related to the superconducting transitions of LaSn$_3$ ($T_{\rm c}\approx$ 6.2 K \cite{Gambino_1968}) and Sn ($T_{\rm c} \approx$ 3.7 K); both phases being seen in the powder diffraction pattern shown in Fig. \ref{fig:x-ray}. 

\begin{figure}
\begin{center}
\resizebox*{7.5cm}{!}{\includegraphics{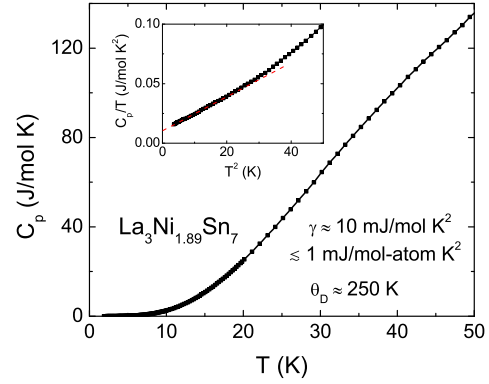}}%
\caption{ Temperature dependence of specific heat of La$_3$Ni$_{1.89}$Sn$_7$ in the form of $C_p (T)$ vs. \textit{T}. Inset: $C_p/T$ vs. $T^2$; dashed line is guide to the eye.}%
\label{fig:La_C}
\end{center}
\end{figure}

Figure \ref{fig:La_C} shows the temperature-dependent specific heat $C_p$ for La$_3$Ni$_{1.89}$Sn$_7$. $C_p$ increases smoothly up to 50 K showing no resolvable features at low temperatures, confirming the anomalies seen in zero-field resistivity are brought by the impurity phases. The electronic specific heat coefficient ($\gamma$) and Debye temperature ($\Theta_D$) were estimated using the relation $C_p/T = \gamma + \beta T^2$ by extrapolating data $C_p/T$ vs. $T^2$ below 7 K (shown in the inset of Fig. \ref{fig:La_C}). The calculated values are  $\gamma \approx$ 10 mJ/mol-formula-unit K$^2$ (or less than 1 mJ/mole-atomic K$^2$), and $\beta \approx$ 1.5 mJ/mol-formula-unit K$^4$, which gives $\Theta_D \approx$ 250 K. 

\subsection{Ce$_3$Ni$_{1.69}$Sn$_7$}

\begin{figure}
\begin{center}
\resizebox*{7.5cm}{!}{\includegraphics{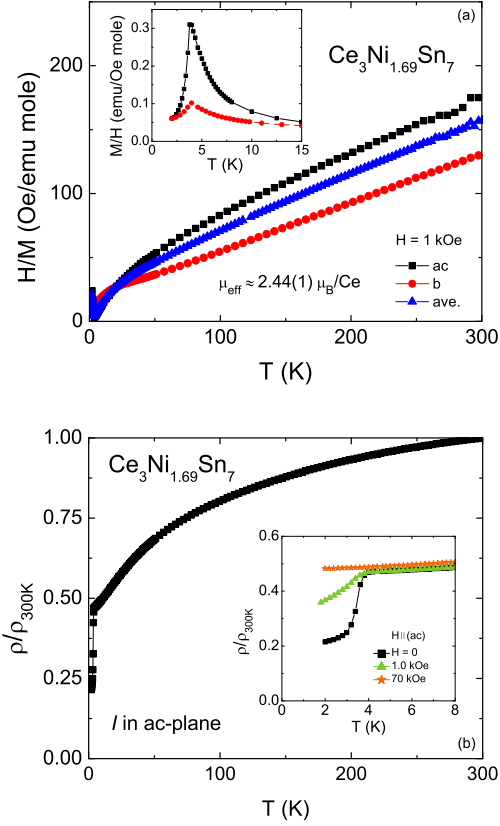}}%
\caption{ (a) Inverse magnetic susceptibility $H/M(T)$ of Ce$_3$Ni$_{1.69}$Sn$_7$ for \textbf{H} $\parallel$ \textit{b}-axis, \textit{ac}-plane and polycrystalline average. Inset: anisotropic magnetic susceptibility below 15 K. (b) Temperature dependence of the normalized electrical resistivity ratio $\rho(T)$/$\rho$(300 K) of Ce$_3$Ni$_{1.69}$Sn$_7$ with $\rho$(300 K) $\sim$ 50 $\mu\Omega$ cm. Inset: low temperature $\rho(T)$/$\rho$(300 K) measured at \textit{H} = 0, 1 and 70 kOe with \textbf{H} $\parallel$ \textit{ac}-plane.}%
\label{fig:Ce_MR}
\end{center}
\end{figure}

The temperature-dependent magnetic susceptibility $\chi(T)=M(T)/H$ and inverse magnetic susceptibility of Ce$_3$Ni$_{1.69}$Sn$_7$ were measured with \textit{H} = 1 kOe applied both parallel to the \textit{b}-axis and the \textit{ac}-plane, and are plotted in Fig. \ref{fig:Ce_MR}(a). The sharp peaks seen at low temperature suggest that this material has an AFM transition, with a larger value of $M(T)/H$ for \textbf{H} $\parallel$ \textit{ac}-plane for $T <$ 15 K. The ordering temperature, consistent with the reported value \cite{SP_2001, Chevalier_1999}, was estimated to be $\sim$ 3.7 K (here and in Table \ref{table:Temperature} the values of the magnetic ordering temperatures obtained from the maximum of the derivatives d($\chi$\textit{T})/d$T$, d$\rho$/d$T$, and/or the specific heat data are quoted). The polycrystalline averaged susceptibility was estimated by $\chi_{ave}$ = $\frac{1}{3}$ $(\chi_{b}+2\chi_{ac})$. The high-temperature magnetic susceptibility can be fitted with the Curie-Weiss law with $\theta_{b}$ = -43.6 K, $\theta_{ac}$ = -75.4 K and $\theta_{ave}$ = -57.1 K. The inferred effective moment from the polycrystalline averaged data: $\mu_{eff}$ = 2.44(1) $\mu_{\rm B}$/Ce is slightly smaller than the expected Hund's rule (\textit{J} = 5/2) ground-state value, 2.54 $\mu_{\rm B}$, but larger than previously reported, 2.33 $\mu_{\rm B}$/Ce \cite{Chevalier_1999}. It should be noted that the anisotropy changes its sign upon cooling in the paramagnetic state ( as can be seen by comparing Fig. \ref{fig:Ce_MR}(a) inset to main body of \ref{fig:Ce_MR}(a)).

The temperature-dependence of the normalized electrical resistivity ratio $\rho(T)$/$\rho$(300 K) for Ce$_3$Ni$_{1.69}$Sn$_7$ is shown in Fig. \ref{fig:Ce_MR}(b). To within a factor of 50\%, the room temperature resistivity $\rho (300 {\rm K})$ reaches approximately 70 $\mu \Omega$ cm, with (RRR) $\simeq$ 3.0. A broad feature is found at around 60 K, which is probably associated with thermal population of the CEF levels. A dramatic drop in the zero-field resistivity value at $\sim$ 3.7 K can be attributed to the near simultaneous occurrence of the $T_{\rm c}$ of the minority Sn phase and a bulk AFM transition. The inset of Fig. \ref{fig:Ce_MR}(b) shows $\rho(T)$/$\rho$(300 K) below 8 K measured at zero field, 1 kOe and 70 kOe for field parallel to the \textit{ac}-plane. The sharp drop in resistivity below 4 K is due, in part, to traces of Sn, but as shown in Fig. \ref{fig:Ce_MR}(b) $T_{\rm N}$ $\sim$ 3.7 K as well. The 1 kOe data show the decrease in resistivity is smaller than in zero field, however, the transition temperature does not change significantly. For the two possible secondary phases in this material, whereas Sn has an upper critical field of 305 Oe at 0 K, no magnetic ordering or superconductivity has been observed for CeSn$_3$ down to low temperatures \cite{Shenoy_1970}. Thus, the sharp drop at 3.7 K for \textit{H} = 1 kOe is primarily caused by the loss of spin disorder scattering. In a higher applied field, 70 kOe, the magnetic ordering has been completely suppressed and no anomaly can be seen (see discussion of metamagnetism below).

\begin{figure}
\begin{center}
\resizebox*{7.5cm}{!}{\includegraphics{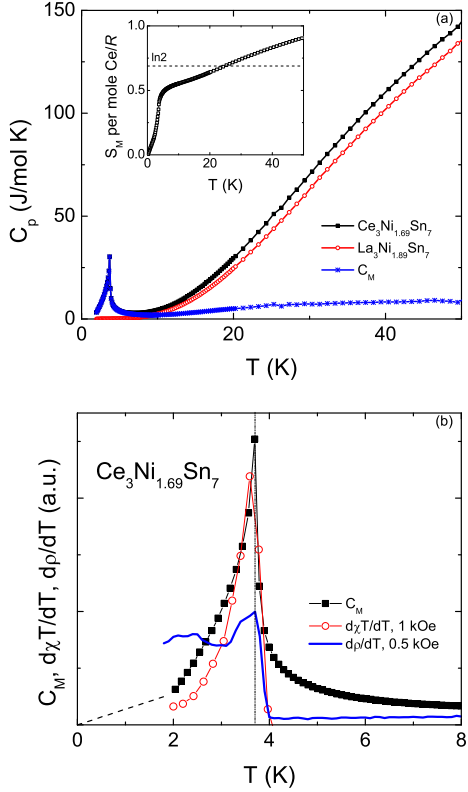}}%
\caption{(a) Specific heat of Ce$_3$Ni$_{1.69}$Sn$_7$ and La$_3$Ni$_{1.89}$Sn$_7$ single crystals and the magnetic specific heat of Ce$_3$Ni$_2$Sn$_7$. Inset: magnetic entropy per mole Ce ion divided by \textit{R}. (b) Low-temperature d($\chi$\textit{T})/d$T$ for \textit{H} = 1 kOe, d$\rho$/d$T$ for \textit{H} = 0.5 kOe and $C_M(T)$ for Ce$_3$Ni$_{1.69}$Sn$_7$. The dashed line indicates $C_M(T)$ extrapolated to \textit{T} = 0. The antiferromagnetic ordering temperature marked by dotted line shows up as a sharp, well-defined peak in all three data sets.}%
\label{fig:Ce_C}
\end{center}
\end{figure}

The specific heat data manifest a sharp rise with decreasing temperature below 6 K, which peaks at $\sim$ 3.7 K (Fig. \ref{fig:Ce_C}(a)). The AFM ordering temperature can be clearly determined as shown in Fig. \ref{fig:Ce_C}(b), which displays the d($\chi$\textit{T})/d$T$, d$\rho$/d$T$ (to suppress Sn's superconducting feature, data of \textit{H} = 0.5 kOe was used), and $C_M(T)$ curves. Each of these data sets gives $T_{\rm N}$ = 3.7 $\pm$ 0.1 K.

Due to the AFM ordering and the broad feature associated with the CEF splitting at higher temperatures, $\gamma$ and $\Theta_D$ of Ce$_3$Ni$_{1.69}$Sn$_7$ cannot be estimated by the same method used with La$_3$Ni$_{1.89}$Sn$_7$. On the other hand, though, the magnetic contribution to specific heat from the Ce ions was calculated by the relation of $C_M$ = $C_p$ (Ce$_3$Ni$_{1.69}$Sn$_7$) - $C_p$ (La$_3$Ni$_{1.89}$Sn$_7$). $C_M$ data show a broad maximum centered around 45 K, indicating a significant magnetic contribution from the Ce ions above $T_{\rm N}$. This broad peak is likely brought by an electric Schottky contribution due to the CEF splitting of the Hund's rule ground state multiplet. The magnetic entropy per mole Ce ion is shown in the inset of Fig. \ref{fig:Ce_C}(a). The $S_M$ reaches about 60$\%$ of \textit{R}ln(2) at $T_{\rm N}$ and recovers the full doublet entropy, \textit{R}ln(2), by 25 K. This might be caused by the Kondo screening of Ce magnetism, or as suggested by a previous neutron study \cite{SP_2001}, not all the Ce ions are participating in the AFM ordering.

\begin{figure}
\begin{center}
\resizebox*{7.5cm}{!}{\includegraphics{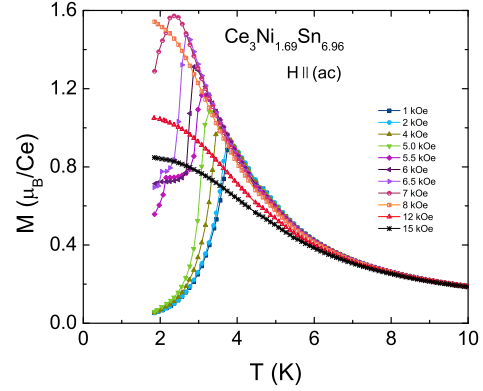}}%
\caption{ $M(T)$ of Ce$_3$Ni$_{1.69}$Sn$_7$ for \textbf{H} $\parallel$ \textit{ac}-plane in selected magnetic fields.}%
\label{fig:Ce_MT}
\end{center}
\end{figure}

The measurements of the low-temperature $M(T)$ with \textbf{H} $\parallel$ \textit{ac}-plane for various applied fields are plotted in Fig. \ref{fig:Ce_MT}. With increasing magnetic field AFM transition systematically shifts to lower temperatures and eventually drops below 2 K for $H >$ 8 kOe. For \textit{H} = 5.5 kOe and 6.5 kOe, another feature emerges at low temperatures, shown as a cusp at $\sim$ 2 K. However, the origin of this feature and its absence at 6.0 kOe are currently unknown. At higher fields, $H >$ 8 kOe, $M(T)$ does not reveal any signature of a phase transition and instead displays a tendency toward saturation at low temperatures. 

\begin{figure}
\begin{center}
\resizebox*{7.5cm}{!}{\includegraphics{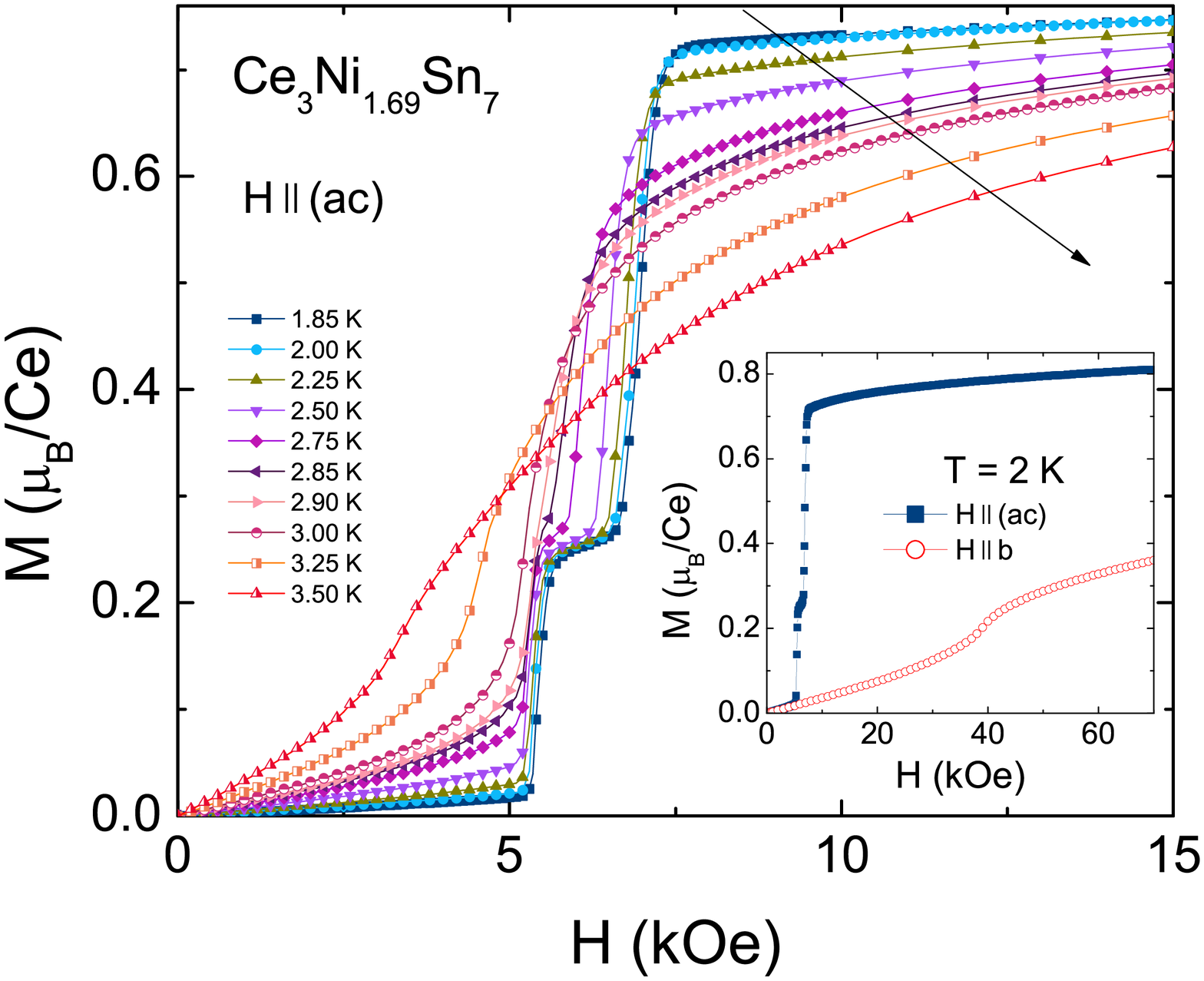}}%
\caption{Magnetization isotherms of Ce$_3$Ni$_{1.69}$Sn$_7$ for \textbf{H} $\parallel(ac)$ at \textit{T} = 1.85, 2.00, 2.25, 2.50, 2.50, 2.75, 2.85, 2.90, 3.00, 3.25 and 3.50 K. The arrow indicates the direction of increasing temperature. Inset: anisotropic field-dependent magnetization of Ce$_3$Ni$_{1.69}$Sn$_7$ at 2 K.}%
\label{fig:Ce_MH}
\end{center}
\end{figure}

\begin{figure}
\begin{center}
\resizebox*{7.5cm}{!}{\includegraphics{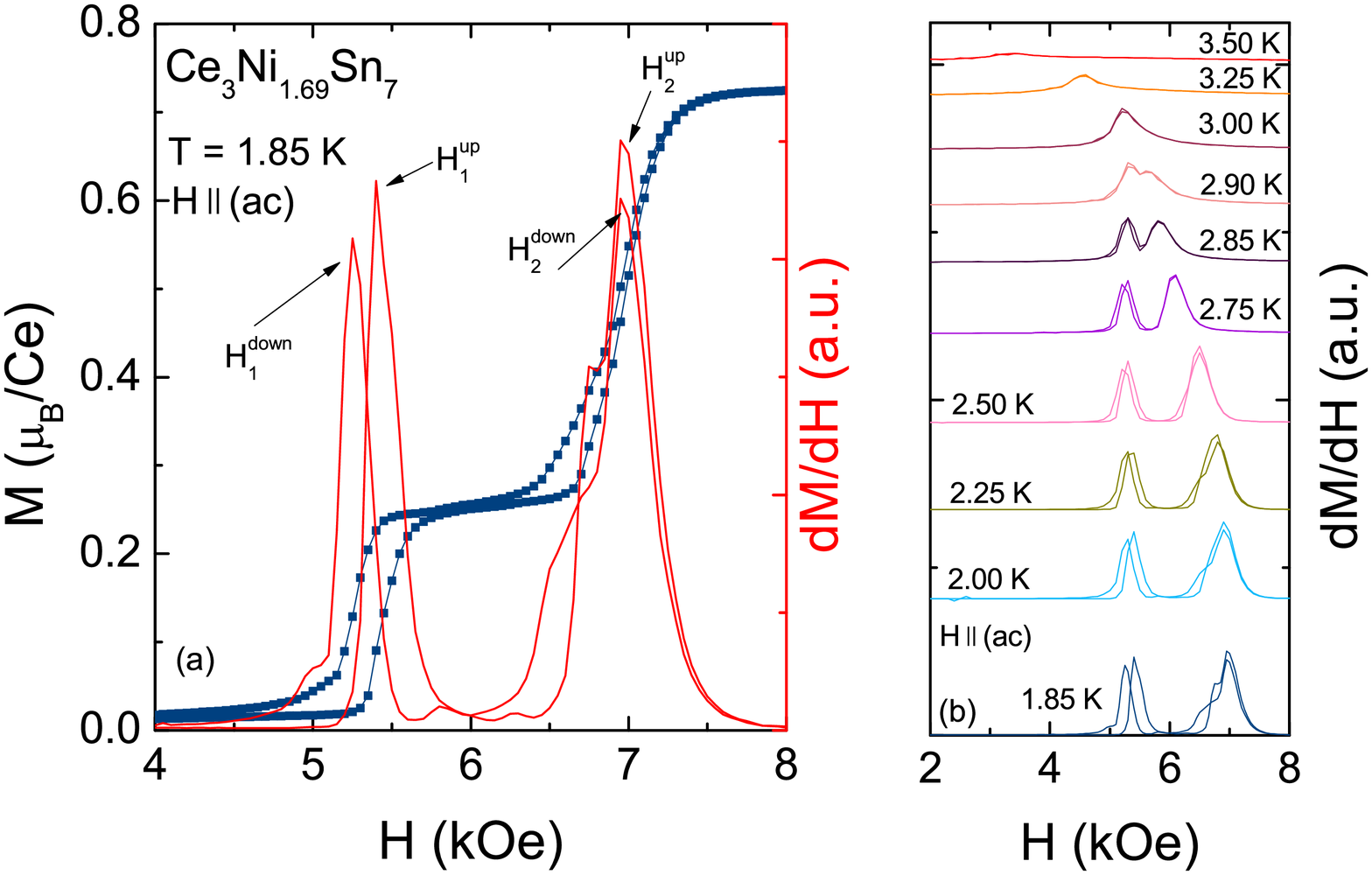}}%
\caption{(a) Magnetization hysteresis data and $dM(H)/dH$ for \textbf{H} $\parallel(ac)$ at \textit{T} = 1.85 K. $H_{1}^{\rm down}$, $H_{1}^{\rm up}$, $H_{2}^{\rm down}$ and $H_{2}^{\rm up}$ indicate lower-field metamagnetic transition measured in decreasing fields, lower-field metamagnetic transition measured in increasing fields, higher-field metamagnetic transition measured in decreasing fields and higher-field metamagnetic transition measured in increasing, respectively. (b) $dM(H)/dH$ for \textbf{H} $\parallel(ac)$ at \textit{T} = 1.85, 2.00, 2.25, 2.50, 2.50, 2.75, 2.85, 2.90, 3.00, 3.25 and 3.50 K.}%
\label{fig:Ce_derivative}
\end{center}
\end{figure}

The anisotropic $M(H)$ isotherms of Ce$_3$Ni$_{1.69}$Sn$_7$ are plotted in the inset of Fig. \ref{fig:Ce_MH}. The observed curves show significant anisotropic behavior at 2 K. For \textbf{H} $\parallel$ \textit{b}-axis, $M(H)$ linearly increases with field up to 40 kOe, followed by a broad metamagnetic transition, then linearly rises to about 0.36 $\mu_{\rm B}$ per Ce ion near 70 kOe. On the other hand, for \textbf{H} $\parallel$ \textit{ac}-plane, at least two metamagnetic transitions take place below 10 kOe, which can be clearly seen in the main body of Fig. \ref{fig:Ce_MH} and also indicated in the $dM(H)/dH$ analysis in Fig. \ref{fig:Ce_derivative}(a). In higher fields, $M(H)$ with \textbf{H} $\parallel$ \textit{ac}-plane linearly approaches to 0.81 $\mu_{\rm B}$ per Ce near 70 kOe, which is well below the expected full moment of 2.14 $\mu_{\rm B}$/Ce. 

Figure \ref{fig:Ce_MH} presents the temperature-dependent evolution of the metamagnetic transitions for \textbf{H} $\parallel$ \textit{ac}-plane. As temperature increased, both metamagnetic transitions were gradually broadened and eventually smeared out at 3.5 K. An examination of the hysteresis associated with metamagnetic transitions is shown in Fig. \ref{fig:Ce_derivative}(a), $M(H)$ at 1.85 K and  $dM(H)/dH$ are plotted for Ce$_3$Ni$_{1.69}$Sn$_7$. The two metamagnetic transitions manifest as two distinct steps, and the hysteresis can been clearly resolved in $M(H)$. Correspondingly, the derivatives reveal the metamagnetic transitions at lower and higher field (indicated as 1 and 2 in the subscript, respectively) during the process of increasing and decreasing magnetic fields (indicated as up and down in the superscript, respectively): $H_{1}^{\rm down}$ = 5.25 kOe, $H_{1}^{\rm up}$ = 5.41 kOe, $H_{2}^{\rm down}$ = 6.95 kOe and $H_{2}^{\rm up}$ = 6.96 kOe (Fig. \ref{fig:Ce_derivative}(a)). Similarly, the derivatives $dM(H)/dH$ at different temperatures provide the temperature evolution of metamagnetic transitions in Fig. \ref{fig:Ce_derivative}(b). As temperature increases, the two metamagnetic transition peaks systematically shift to lower fields and broaden. At 3.0 K, the two metamagnetic transitions merge into one and eventually vanishes at $\sim$ 3.5 K.

\begin{figure}
\begin{center}
\resizebox*{7.5cm}{!}{\includegraphics{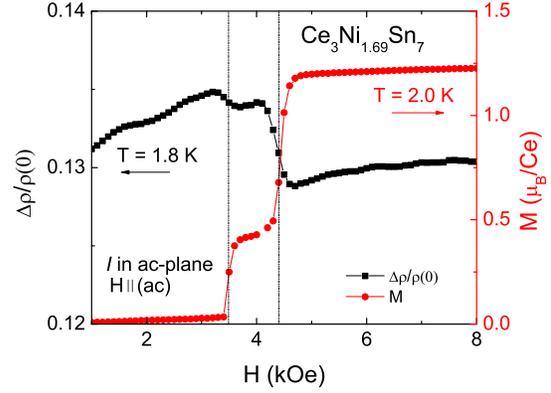}}%
\caption{ Magnetoresistance (left axis) as a function of magnetic field for \textit{T} = 1.8 K and magnetization (right axis) as a function of magnetic field at 2.0 K for \textbf{H} $\parallel$ \textit{ac}-plane. Note: both measurements were done on the same sample in the same orientation; dashed lines indicate the magnetic fields where $dM/dH$ peaks.}%
\label{fig:Ce_RMH}
\end{center}
\end{figure}

In order to correlate features in $\rho(H)$ and $M(H)$, as well as establish in-plane anisotropic response to applied field \cite{Myers_1999, Canfield_1997}, the field dependence of mangetoresistance at \textit{T} = 1.8 K and magnetization at \textit{T} = 2.0 K are shown together in Fig. \ref{fig:Ce_RMH} from measurements on the same sample in the same in-plane orientation of \textbf{H} $\parallel$ \textit{ac}. Two metamagnetic transitions can be clearly seen and are in  good agreement when derived from the magnetoresistance and magnetization curves. The observed metamagnetic transitions fields are lower than previously discussed (see Fig. \ref{fig:Ce_MH}), this is possibly due to the in-plane anisotropy \cite{Myers_1999, Canfield_1997}.  

\begin{figure}
\begin{center}
\resizebox*{7.5cm}{!}{\includegraphics{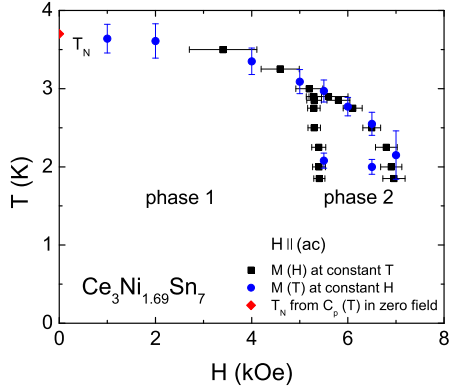}}%
\caption{$H-T$ phase diagram of Ce$_3$Ni$_{1.69}$Sn$_7$ for \textbf{H} $\parallel(ac)$, measured on the same sample in the same orientation. The transition points are taken from the derivatives of $M(H)$ at constant temperatures in the process of increasing fields sweep and the derivatives of $M(T)$ at constant fields in the process of increasing temperature. The error bars are taken as the full width at half maximum of the derivatives.}%
\label{fig:Ce_HT}
\end{center}
\end{figure}

Based on the above discussion, an $H-T$ phase diagram is constructed (presented in Fig. \ref{fig:Ce_HT}), where the metamagnetic transition points are extracted from the temperature dependent magnetic susceptibilities and magnetization isotherms for one, arbitrary field orientation in the \textit{ac}-plane. From the data presented in Figs. \ref{fig:Ce_MT} and \ref{fig:Ce_MH}, for 0 $< H \lesssim$ 5 kOe $T_{\rm N}$ is gradually suppressed from $T_{\rm N} \approx $ 3.7 K at \textit{H} = 0 to $T_{\rm N} \approx $ 3.1 K at \textit{H} = 5 kOe. For 5 kOe $\lesssim H \lesssim$ 7.2 kOe a second line in the $H-T$ phase diagram appears. Where as the outer envelop of the $H-T$ diagram continues to show a gradual, but non-linear suppression of $T_{\rm N}$ with \textbf{H}, a second, near vertical line appears for $H \sim$ 5.2 kOe.

Figure. \ref{fig:Ce_HT}, taken together with the implicit in \textit{ac}-plane anisotropy suggested by comparison to Fig. \ref{fig:Ce_RMH}, make it clear that Ce$_3$Ni$_{1.69}$Sn$_7$ will have a rich $M (T, H, \theta)$ ($\theta$ being in plane angle of field with respect to \textit{a}-axis) phase diagram.

\subsection{Pr$_3$Ni$_{1.56}$Sn$_7$}

\begin{figure}
\begin{center}
\resizebox*{7.5cm}{!}{\includegraphics{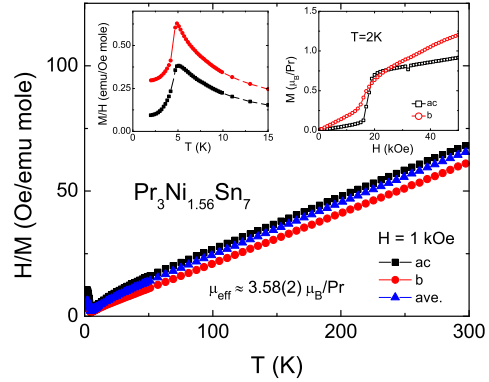}}%
\caption{Inverse magnetic susceptibility $H/M(T)$ of Pr$_3$Ni$_{1.56}$Sn$_7$ for \textbf{H} $\parallel$ \textit{b}-axis, in \textit{ac}-plane and polycrystalline average at $H$ = 1 kOe. Left inset: enlarged anisotropic magnetic susceptibility below 15 K. Right inset: anisotropic $M(H)$ of Pr$_3$Ni$_{1.56}$Sn$_7$ at \textit{T} = 2 K.}%
\label{fig:Pr_MT}
\end{center}
\end{figure}

The anisotropic magnetic properties of Pr$_3$Ni$_{1.56}$Sn$_7$ are shown in Fig. \ref{fig:Pr_MT}. As revealed by the inverse magnetic susceptibility measured at 1 kOe, the magnetic susceptibility follows the Curie-Weiss law at high temperatures, $\theta_{b}$ = -5.2 K, $\theta_{ac}$ = -26.7 K and $\theta_{ave}$ = -17.7 K. This anisotropy in the paramagnetic state results in $(M/H)_{b} > (M/H)_{ac}$ over the whole temperature range measured. The effective moment obtained from the fit of polycrystalline average susceptibility is $\mu_{eff}$ = 3.58(2) $\mu_{\rm B}$ per Pr$^{3+}$ ion (see Table \ref{table:Temperature}), identical to the free ion value for Pr$^{3+}$. In the low temperature region (shown in the left inset of Fig. \ref{fig:Pr_MT}), sharp peaks in $\chi(T)$, at \textit{T} $\sim$ 4.8 K, indicate AFM transition. The anisotropic field-dependent magnetization isotherms of Pr$_3$Ni$_{1.56}$Sn$_7$ measured at 2 K are shown in the right inset of Fig. \ref{fig:Pr_MT}. For both orientations, $M(H)$ linearly increases as the applied field increases, followed by a broad metamagnetic transition occurring at $\sim$ 16 kOe for \textbf{H} $\parallel$ \textit{b} and $\sim$ 17.5 kOe for \textbf{H} $\parallel$ (\textit{ac}). Since up to 50 kOe $M(H)$ for both orientations does not show saturation and the values of the magnetization at 50 kOe are much lower than expected for Pr$^{3+}$ (3.2 $\mu_{\rm B}$), it is likely that in higher fields more metamagnetic transitions will occur. 

\begin{figure}
\begin{center}
\resizebox*{7.5cm}{!}{\includegraphics{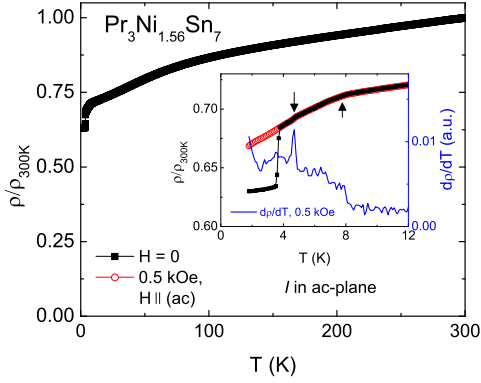}}%
\caption{Temperature dependence of the normalized electrical resistivity ratio $\rho(T)$/$\rho$(300 K) of Pr$_3$Ni$_{1.56}$Sn$_7$ with $\rho$(300 K) $\sim$ 60 $\mu\Omega$ cm. Inset: low temperature $\rho(T)$/$\rho$(300 K) (left axis) measured at \textit{H} = 0, 0.5 kOe with \textbf{H} $\parallel$(\textit{ac}) and d$\rho$/d$T$ (right axis) at \textit{H} = 0.5 kOe. Arrows indicate the transition temperature 4.7 and 7.8 K, respectively.}%
\label{fig:Pr_RT}
\end{center}
\end{figure}

The temperature-dependence of the normalized electrical resistivity ratio for Pr$_3$Ni$_{1.56}$Sn$_7$ is shown in Fig. \ref{fig:Pr_RT}. To within factor a of 50\%, the room temperature resistivity $\rho$(300 K) reaches approximately 60 $\mu \Omega$ cm, with (RRR) $\simeq$ 1.5. Resistivity decreases with decreasing temperature and shows a broad feature at around 50 K, which can be attributed to the thermal population of CEF levels. The enlarged low-T resistivity in zero field reveals three successive anomalies (see inset of Fig. \ref{fig:Pr_RT}), occurring at 3.7 K, 4.7 K and 7.8 K. The sharp transition at 3.7 K is suppressed by \textit{H} = 0.5 kOe making it probable that it is due to a small amount of residual Sn. The 4.7 K feature is seen as a subtle drop in resistivity, but manifests itself as a sharp peak in $d\rho/dT$. This feature coincides with the AFM transition seen in the magnetic susceptibility and is associated with the loss of the spin-disorder scattering. Although difficult to see in the resistivity data, d$\rho$/d$T$ shows a clear step-like anomaly at $\simeq$ 6.8 K. At 7.8 K, the resistivity changes its slope, and exhibits a step-like feature in d$\rho$/d$T$ as well. 

\begin{figure}
\begin{center}
\resizebox*{7.5cm}{!}{\includegraphics{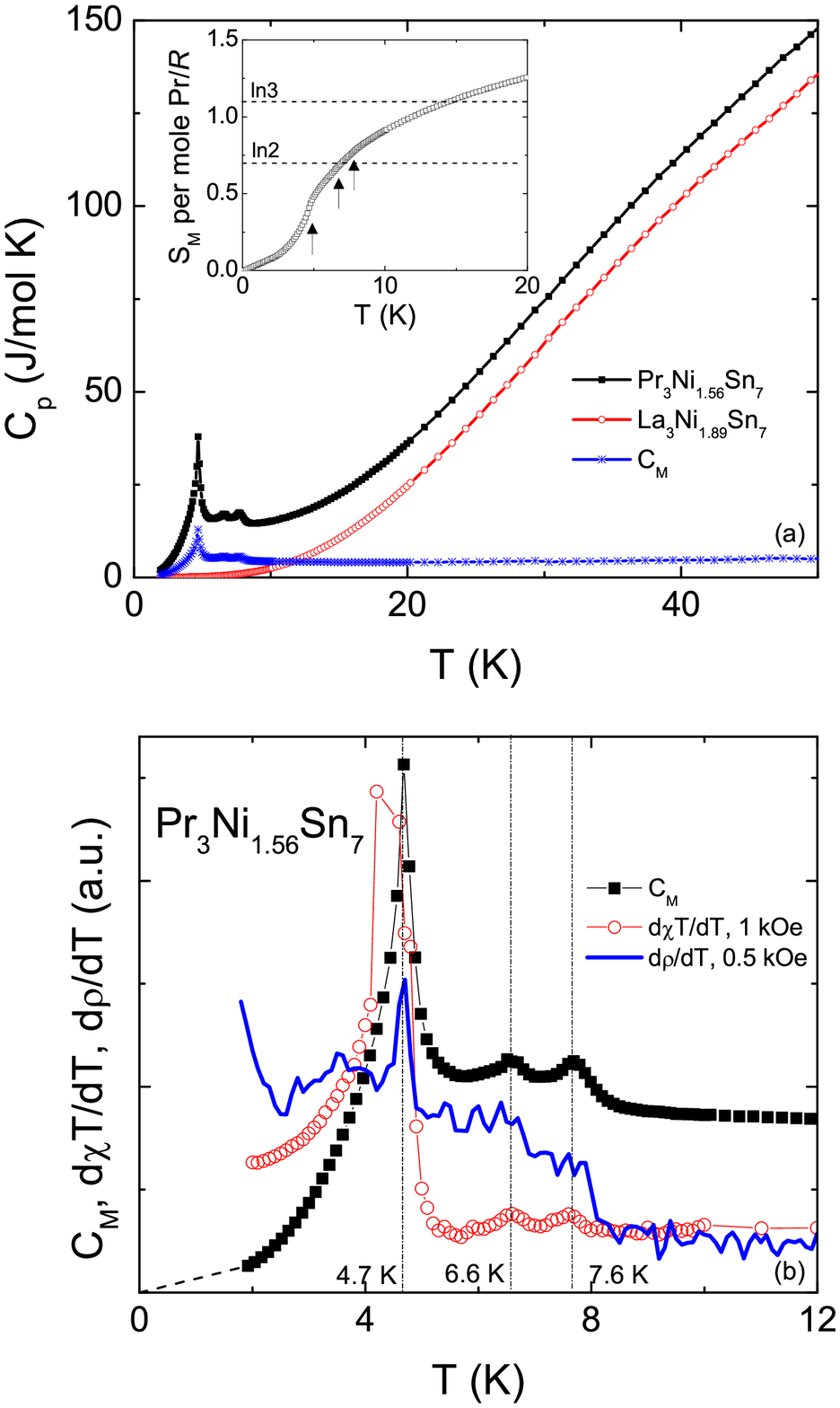}}%
\caption{(a) Specific heat of Pr$_3$Ni$_{1.56}$Sn$_7$ and La$_3$Ni$_2$Sn$_7$ single crystals and the magnetic specific heat of Pr$_3$Ni$_{1.56}$Sn$_7$. Inset: magnetic entropy per Pr$^{3+}$ ion divided by \textit{R}. The arrows indicate the transition temperatures. (b) Low-temperature d($\chi$\textit{T})/d$T$ for \textit{H} = 1 kOe, d$\rho$/d$T$ for \textit{H} = 0.5 kOe and $C_M(T)$ for Pr$_3$Ni$_{1.56}$Sn$_7$. The dashed line indicates $C_M(T)$ extrapolated to \textit{T} = 0. The dotted lines mark the transitions in all three plots.}%
\label{fig:Pr_C}
\end{center}
\end{figure}

Specific heat of Pr$_3$Ni$_{1.56}$Sn$_7$ initially decreases with decreasing temperature, and reveals three anomalies, which peak at $\sim$ 7.6 K, 6.6 K and 4.7 K (Fig. \ref{fig:Pr_C}(a)). Enlarged low temperature data of d($\chi$\textit{T})/d$T$, d$\rho$/d$T$ and $C_M(T)$ are shown in Fig. \ref{fig:Pr_C}(b). It is clear that two more anomalies are seen in the derivative d($\chi$\textit{T})/d$T$, which corroborates the specific heat results very well. d$\rho$/d$T$ exhibits a peak at 4.7 K and two step-like features at 6.8 and 7.6 K. Based on the results of thermodynamic and transport measurements, for Pr$_3$Ni$_{1.56}$Sn$_7$, the magnetic ordering temperatures are 7.6 K, 6.6 K and 4.7 K. The magnetic contribution to specific heat from Pr$^{+3}$ ions was calculated by the relation of $C_M$ = $C_p$ (Pr$_3$Ni$_{1.56}$Sn$_7$) - $C_p$ (La$_3$Ni$_{1.89}$Sn$_7$). The magnetic entropy $S_M$ per mole Pr$^{3+}$ (shown in the inset of Fig. \ref{fig:Pr_C}(a)) is roughly about \textit{R}ln(2) at $T$ = 6.8 K, and becomes \textit{R}ln(3) by 14.3 K. 

\begin{figure}
\begin{center}
\resizebox*{7.5cm}{!}{\includegraphics{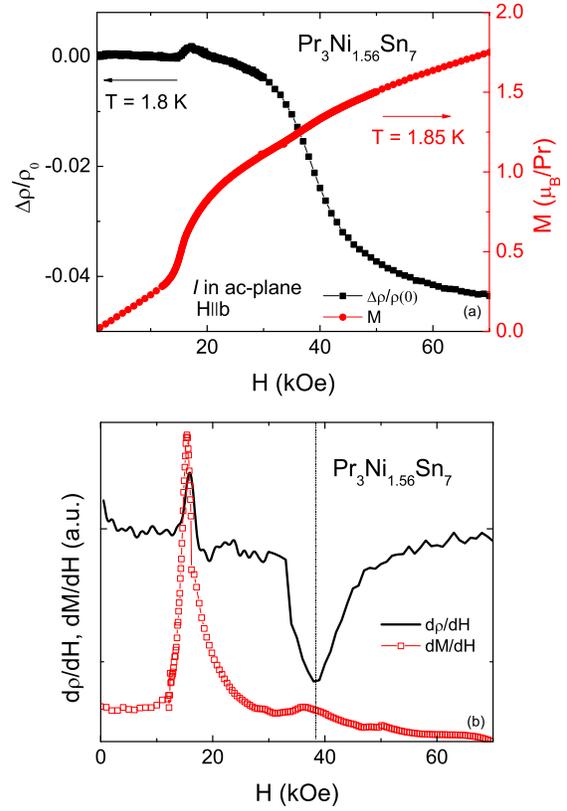}}%
\caption{(a) Magnetoresistance (left axis) as a function of magnetic field for \textit{H} at $T$ = 1.8 K and magnetization (right axis) as a function of magnetic field for \textit{H} at $ T$ = 1.85 K for \textbf{H} $\parallel$ \textit{b}. Note: magnetoresistance was normalized to $\rho$(0.5 kOe) to remove Sn off-set. (b) $d\rho/dH$ at $T$ = 1.8 K and $dM/dH$ at $ T$ = 1.85 K for \textbf{H} $\parallel$ \textit{b}. Note: both measurements were done on the same sample in the same orientation.}%
\label{fig:Pr_RH}
\end{center}
\end{figure}

The field dependence of the magnetoresistance and magnetization for Pr$_3$Ni$_{1.56}$Sn$_7$ are shown in Fig. \ref{fig:Pr_RH}(a). To get rid of an off-set associated with superconducting Sn, the magnetoresistance was normalized to $\rho$(0.5 kOe), i.e. $\bigtriangleup \rho$/$\rho_0$ = ($\rho(H)$ - $\rho$(0.5 kOe))/$\rho$(0.5 kOe).  There are two clear metamagnetic transitions visible in the 1.8 K data, one at $\sim$ 17 kOe and a second one near 38 kOe. Although both are clearly seen in the magnetoresistance data, the higher field transition is more clearly seen magnetization via $dM/dH$ plots (Fig. \ref{fig:Pr_RH}(b)).

\subsection{Nd$_3$Ni$_{1.34}$Sn$_7$}

\begin{figure}
\begin{center}
\resizebox*{7.5cm}{!}{\includegraphics{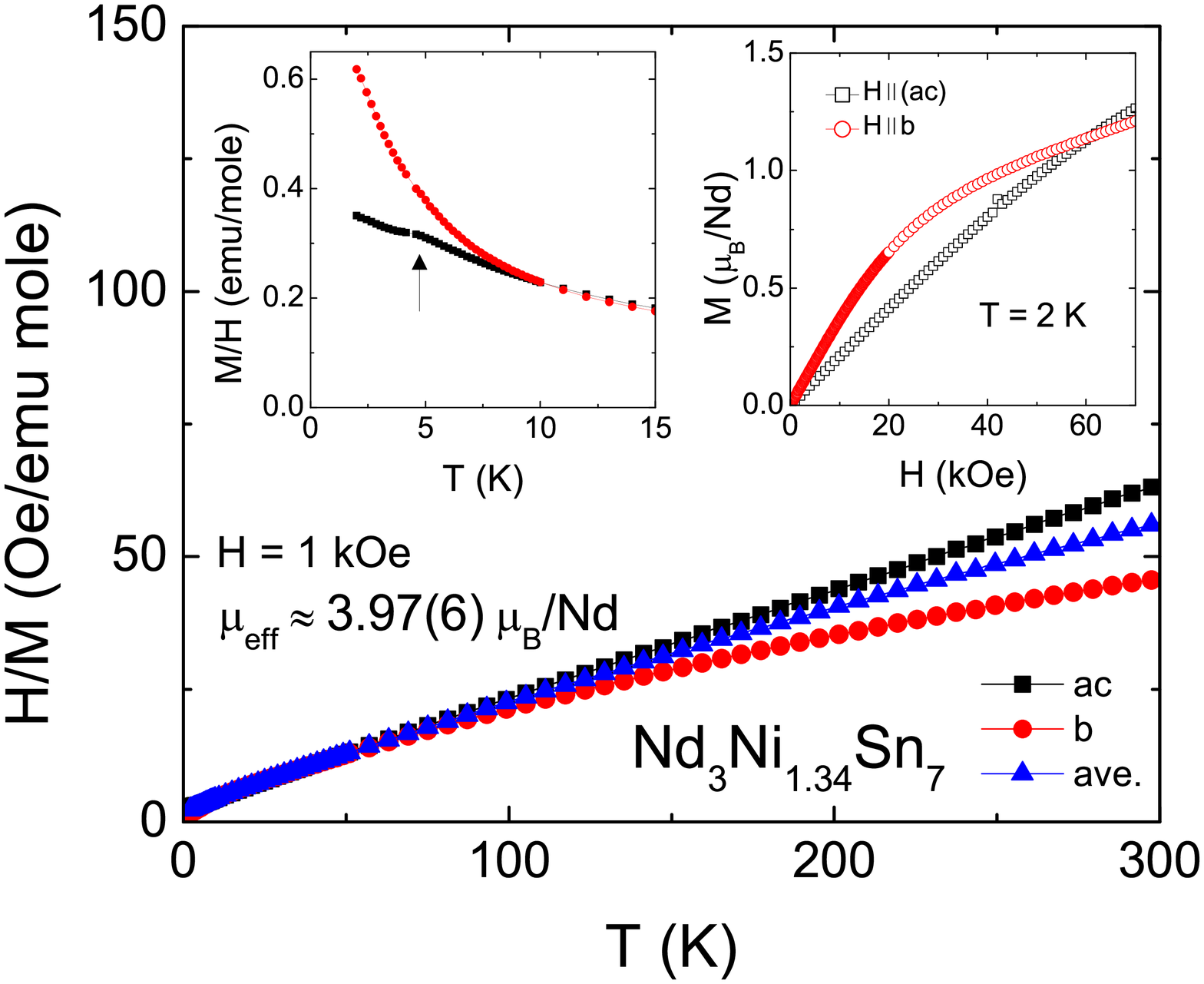}}%
\caption{Inverse magnetic susceptibility $H/M(T)$ of Nd$_3$Ni$_{1.34}$Sn$_7$ for \textbf{H} $\parallel$ \textit{b}-axis and \textit{ac}-plane and polycrystalline average at $H$ = 1 kOe. Left inset: enlarged anisotropic magnetic susceptibility below 15 K. Right inset: anisotropic $M(H)$ of Nd$_3$Ni$_{1.34}$Sn$_7$ at \textit{T} = 2 K.}%
\label{fig:Nd_M}
\end{center}
\end{figure}

For Nd$_3$Ni$_{1.34}$Sn$_7$, magnetization with the applied magnetic field $H$ = 1 kOe is found to be anisotropic with $(M/H)_{b}t > (M/H)_{ac}$ (Fig. \ref{fig:Nd_M}). At high temperatures, the magnetic susceptibility follows the Curie-Weiss law, resulting in $\theta_{ave}$ = -36.8 K and $\mu_{eff}$ = 3.97(6) $\mu_{\rm B}$ per Nd$^{3+}$ ion (see Table \ref{table:Temperature}), slightly larger than 3.87 $\mu_{\rm B}$, the expected value for Nd$^{3+}$ free ion. At low temperatures, Nd$_3$Ni$_{1.34}$Sn$_7$ enters antiferromagnetic state at $\sim$ 3.8 K, seen by a subtle cusp in the magnetic susceptibility curve (left inset of Fig. \ref{fig:Nd_M}). Magnetization isotherms of Nd$_3$Ni$_{1.34}$Sn$_7$ measured at 2 K are provided in the right inset of Fig. \ref{fig:Nd_M}.

\begin{figure}
\begin{center}
\resizebox*{7.5cm}{!}{\includegraphics{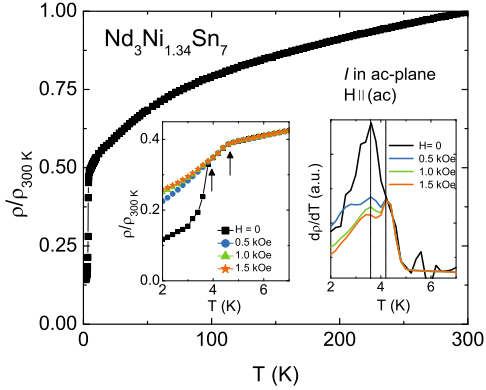}}%
\caption{Temperature dependence of the normalized electrical resistivity ratio $\rho(T)$/$\rho$(300 K) of Nd$_3$Ni$_{1.34}$Sn$_7$ with $\rho$(300 K) $\sim$ 100 $\mu\Omega$ cm. Left inset: low temperature $\rho(T)$/$\rho$(300 K) (left axis) measured at \textit{H} = 0, 0.5 kOe, 1.0 kOe and 1.5 kOe with \textbf{H} $\parallel$(\textit{ac}). Arrows indicate the anomalies at 3.8 K and 4.6 K. Right Inset: d$\rho$/d$T$ at \textit{H} = 0, 0.5 kOe, 1.0 kOe and 1.5 kOe.}%
\label{fig:Nd_R}
\end{center}
\end{figure}

The temperature dependence of the normalized electrical resistivity ratio for Nd$_3$Ni$_{1.34}$Sn$_7$ is shown in Fig. \ref{fig:Nd_R}. To within a factor of 50\%, the room temperature resistivity $\rho$ (300 K) reaches approximately 100 $\mu \Omega$ cm, with (RRR) $\simeq$ 4.4 in zero magnetic field. A similar broad feature is seen at the higher temperatures and implies the thermal population of CEF levels. The enlarged low-T resistivity ratio measured at several selected magnetic fields is plotted in the left inset of Fig. \ref{fig:Nd_R}. The higher-temperature feature, the break in slope of the resistivity occurring at 4.6 K, does not shift with different magnetic fields. A second anomaly at $\sim$ 3.8 K in zero field almost disappears when measured with applied fields. This indicates the anomaly is possibly associated with the residual Sn. However, in $d\rho/dT$ (see right inset of Fig. \ref{fig:Nd_R}), the cusp at $\sim$ 3.6 K, does not disappear or shift with applied fields. Thus, it is likely that Nd$_3$Ni$_{1.34}$Sn$_7$ has a second, lower-temperature magnetic transition at $T_{\rm N}$ $\sim$ 3.6 K, overlapping with the $T_{\rm c}$ ($H$ = 0 ) of Sn. 

\begin{figure}
\begin{center}
\resizebox*{7.5cm}{!}{\includegraphics{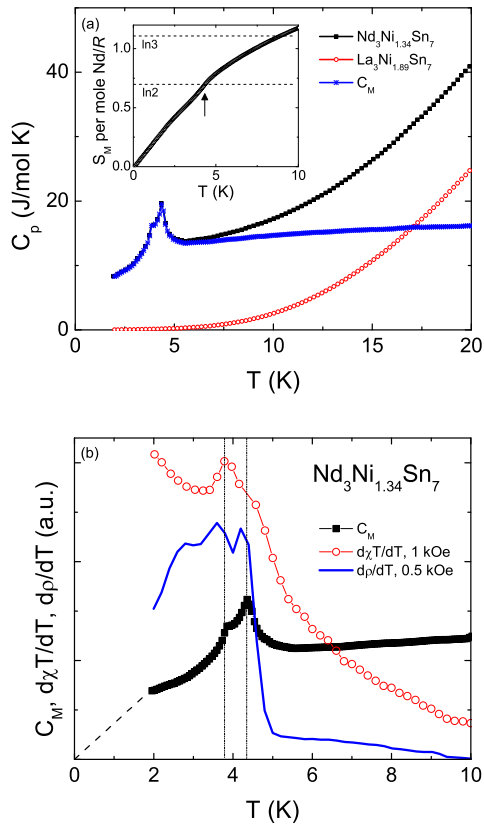}}%
\caption{(a) Specific heat of Nd$_3$Ni$_{1.34}$Sn$_7$ and La$_3$Ni$_{1.89}$Sn$_7$ single crystals and the magnetic specific heat of Nd$_3$Ni$_{1.34}$Sn$_7$. Inset: magnetic entropy per mole Nd$^{3+}$ ion divided by \textit{R}. The arrow indicates the anomaly at \textit{T} = 4.3 K. (b) Low-temperature d($\chi$\textit{T})/d$T$ for \textit{H} = 1 kOe, d$\rho$/d$T$ for \textit{H} = 0.5 kOe and $C_M(T)$ for Nd$_3$Ni$_{1.34}$Sn$_7$. The dashed line indicates $C_M(T)$ extrapolated to \textit{T} = 0. The dotted lines mark the transitions in all three plots.}%
\label{fig:Nd_C}
\end{center}
\end{figure}

The specific heat data of Nd$_3$Ni$_{1.34}$Sn$_7$ are shown in Fig. \ref{fig:Nd_C} (a). Two anomalies are observed at 3.8 K and 4.3 K. Enlarged low temperature data of d($\chi$\textit{T})/d$T$, d$\rho$/d$T$ and $C_M(T)$ are shown in Fig. \ref{fig:Nd_C}(b). The transition at 3.8 K coincides in both magnetic susceptibility and specific heat data, the corresponding transition shifts to 3.6 K in transport measurement. This sharp feature in the specific heat data further indicates Nd$_3$Ni$_{1.34}$Sn$_7$ has a transition at $\sim$ 3.8 K. The higher-temperature (4.3 K) anomaly in the specific heat seems to find its counterpart at \textit{T} = 4.2 K in the transport data. There is a subtle change of change of slope in the similar temperature region seen in d($\chi$\textit{T})/d$T$. The magnetic contribution to specific heat from Nd$^{+3}$ ions was calculated by the relation of $C_M$ = $C_p$ (Nd$_3$Ni$_{1.34}$Sn$_7$) - $C_p$ (La$_3$Ni$_{1.89}$Sn$_7$). The magnetic entropy $S_M$ per mole Nd$^{3+}$ (shown in the inset of Fig. \ref{fig:Nd_C}(a)) reaches approximately \textit{R}ln(2) at $T$ = 4.3 K, the full doublet entropy.

\section{Summary and conclusions}

Motivated by previous studies of rare-earth compounds \cite{Sergey_1999, Sefat_2008, Mun_2010}, we have synthesized single crystalline \textit{R}$_3$Ni$_{2-x}$Sn$_7$ (\textit{R} = La, Ce, Pr and Nd) samples via self-flux Sn. Detailed thermodynamic and transport measurements were performed to study the properties of \textit{R}$_3$Ni$_{2-x}$Sn$_7$ series. The crystals form as plates (inset of Fig. \ref{fig:x-ray}), and can be identified as having an orthorhombic La$_3$Co$_2$Sn$_7$-type structure. We have determined the degree of Ni-site vacancy and see clear evidence of the associated, disorder scattering manifest in the low RRR values. Despite the partial Ni site occupancy there are sharp magnetic transitions and metamagnetic transitions. Although rich and complex $H-T$ phase diagrams are likely, the combination of orthorhombicity and partial Ni site occupancy makes this a less than ideal system for detailed studies. 
  
\section*{Acknowledgements}
We thank W. E. Straszheim for his assistance with the elemental analysis of the samples and S. Kim for her critical reading of the text. This work was carried out at the Iowa State University and supported by the AFOSR-MURI grant No. FA9550-09-1-0603 (X. Lin and P. C. Canfield). Part of this work was performed at Ames Laboratory, US DOE, under contract No. DE-AC02-07CH 11358 (S. L. Bud'ko and S. Thimmaiah).

\begin{table*}[ht]
\caption{Refined unit cell parameters refined from powder x-ray diffraction for \textit{R}$_3$Ni$_{2-x}$Sn$_7$ (\textit{R}= La, Ce, Pr and Nd) compounds.} 
\centering
\begin{tabular}{c c c c c}
\hline\hline
\textit{R} & \textit{a} (\r{A})& \textit{b} (\r{A}) & \textit{c} (\r{A}) & \textit{V} ({\r{A}}$^3$) \\ [0.5ex]
\hline
La	& 4.599(16) & 27.549(24) & 4.602(10) & 583(3)\\
Ce	& 4.564(11) & 27.276(76) & 4.558(10) & 567(4)\\
Pr	& 4.547(15) & 27.2162(10) & 4.546(5) & 563(2)\\
Nd	& 4.530(12) & 27.111(49) & 4.529(7) & 556(2)\\
[1ex]
\hline
\end{tabular}
\label{table:powder}
\end{table*}

\begin{table*}[ht]
\caption{Unit cell parameters and Ni site occupancies obtained from single crystal x-ray diffraction for \textit{R}$_3$Ni$_{2-x}$Sn$_7$ (\textit{R}= La, Ce, Pr and Nd).} 
\centering

\begin{tabular}{c c c c c c c}
\hline\hline
\textit{R} & \textit{a} (\r{A})& \textit{b} (\r{A}) & \textit{c} (\r{A}) & \textit{V} ({\r{A}}$^3$) & 2-\textit{x} (Ni) & Stoichiometry (x-ray) \\ [0.5ex]
\hline
La	& 4.6033(6)	& 27.578(4)	&4.6133(6)	&585.66(13) &1.89(1) & La$_3$Ni$_{1.89}$Sn$_{7}$\\
Ce	& 4.5565(15)&27.300(9)	& 4.5720(15) &568.7(3)  &1.69(1) & Ce$_{3}$Ni$_{1.69}$Sn$_{7}$\\
Pr	& 4.5260(16)&27.173(10)	&4.5475(16)	&559.3(3)	&1.56(1) & Pr$_{3}$Ni$_{1.56}$Sn$_{7}$ \\
Nd	&4.5193(17)	&27.091(10) &4.5408(17)	&555.9(4)	&1.34(1) & Nd$_{3}$Ni$_{1.34}$Sn$_{7}$\\
[1ex]
\hline
\end{tabular}

\label{table:unit}
\end{table*}

\begin{table*}[ht]
\caption{Atomic coordinates and isotropic displacement parameters for \textit{R}$_3$Ni$_{2-x}$Sn$_7$ (\textit{R}= La, Ce, Pr and Nd) single crystals.} 
\centering
\setlength{\tabcolsep}{2.5pt}
\begin{tabular}{c c c c c c c} 
\hline\hline
Atom & Site & Occupancy & x & y & z & Ueq \\ [0.5ex]
\hline
La1	&2\textit{c} &1	&0.5	&0	&0.5	&0.09(1)\\
La2	&4\textit{i} &1	&0	&0.1843(1)	&0	&0.09(1)\\
Ni1	&4\textit{j} &0.95(1)&0	&0.3718(1)	&0.5&0.11(1)\\
Sn1	&2\textit{a} &1  &0	&0	&0	&0.15(1)\\
Sn2	&4\textit{i} &1	&0&	0.4102(1)	&0	&0.12(1)\\
Sn3	&4\textit{j} &1	&0	&0.0900(1)	&0.5	&0.12(1)\\
Sn4	&4\textit{j} &1	&0	&0.2813(1)	&0.5	&0.12(1)\\\\

Ce1	&2\textit{c} &1	&0.5	&0	&0.5	&0.10(1)\\
Ce2	&4\textit{i} &1	&0	&0.1846(1)	&0	&0.11(1)\\
Ni1	&4\textit{j} &0.85(1)	&0	&0.3718(1)	&0.5	&0.13(1)\\
Sn1	&2\textit{a} &1	&0	&0	&0	&0.11(1)\\
Sn2	&4\textit{i} &1	&0	&0.4098(1)	&0	&0.14(1)\\
Sn3	&4\textit{j} &1	&0	&0.0902(1)	&0.5	&0.14(1)\\
Sn4	&4\textit{j} &1	&0	&0.2821(1)	&0.5	&0.16(1)\\\\

Pr1	&2\textit{c} &1	&0.5	&0	&0.5	&0.09(1)\\
Pr2	&4\textit{i} &1	&0	&0.1851(1)	&0	&0.10(1)\\
Ni1	&4\textit{j} &0.78(1)	&0	&0.3716(1)	&0.5	&0.13(1)\\
Sn1	&2\textit{a} &1	&0	&0	&0	&0.12(1)\\
Sn2	&4\textit{i} &1	&0	&0.4090(1)	&0	&0.14(1)\\
Sn3	&4\textit{j} &1	&0	&0.0912(1)	&0.5	&0.14(1)\\
Sn4	&4\textit{j} &1	&0	&0.2823(1)	&0.5	&0.18(1)\\\\

Nd1	&2\textit{c} &1	&0.5	&0	&0.5	&0.10(1)\\
Nd2	&4\textit{i} &1	&0	&0.1855(1)	&0	&0.11(1)\\
Ni1	&4\textit{j} &0.67(1)	&0	&0.3715(1)	&0.5	&0.13(1)\\
Sn1	&2\textit{a} &1	&0	&0	&0	&0.14(1)\\
Sn2	&4\textit{i} &1	&0	&0.4084(1)	&0	&0.15(1)\\
Sn3	&4\textit{j} &1	&0	&0.0919(1)	&0.5	&0.16(1)\\
Sn4	&4\textit{j} &1	&0	&0.2830(1)	&0.5	&0.21(1)\\
[1ex]
\hline
\end{tabular}
\label{table:occupancy}

\end{table*}

\begin{table*}[ht]
\caption{WDS elemental analysis (in atomic $\%$) for \textit{R}$_3$Ni$_{2-x}$Sn$_7$ single crystals.} 
\centering
\setlength{\tabcolsep}{2.5pt}
\begin{tabular}{c c c c c} 
\hline\hline 
Compound & \textit{R} & Ni & Sn & Stoichiometry (WDS) \\ [0.5ex]
\hline
La & 23.67	& 15.78 & 57.23 & La$_{3}$Ni$_{2.00}$Sn$_{7.25}$\\
Ce & 25.73	& 14.80 & 59.47 & Ce$_{3}$Ni$_{1.72}$Sn$_{6.93}$\\
Pr & 25.47 & 13.83	& 60.70	& Pr$_{3}$Ni$_{1.63}$Sn$_{7.15}$ \\
Nd & 26.39 & 12.44 & 61.16	& Nd$_{3}$Ni$_{1.41}$Sn$_{6.95}$\\
[1ex]
\hline
\end{tabular}
\label{table:WDS}
\end{table*}

\begin{table*}[ht]
\caption{Magnetic ordering temperatures, anisotropic Curie temperatures and effective magnetic moment in paramagnetic state for \textit{R}$_3$Ni$_{2-x}$Sn$_7$.} 
\centering
\setlength{\tabcolsep}{2.5pt}
\begin{tabular}{c c c c c c} 
\hline\hline 
Compound&$\theta_{b}$ (K)&$\theta_{ac}$ (K)&$\theta_{ave}$ (K)&$\mu_{eff}$ ($\mu_{\rm B}$)&$T_M$ (K)  \\ [0.5ex]
\hline
Ce &-43.6& -75.4 & -57.1& 2.44& 3.7\\
Pr &-5.2& -26.7	& -17.7	&3.58 &7.6, 6.6, 4.7 \\
Nd &-90.8 & -14.6& -36.8& 3.97& 4.3, 3.8\\
[1ex]
\hline
\end{tabular}
\label{table:Temperature}
\end{table*}

\end{document}